\documentclass[twocolumn,showpacs,aps]{revtex4}
\usepackage[latin2]{inputenc}
\usepackage{graphicx}

\newcommand{\newc}{\newcommand}
\def\Rslash{\slash \!\!\!\! R}
\newc{\hc}{{\rm\,h.c.}}
\newc{\GeV}{{\rm\, GeV}}

\newc{\superhu}{\hat H_u}
\newc{\superhd}{\hat H_d}
\newc{\superl}{\hat L}
\newc{\superr}{\hat R}
\newc{\superec}{\hat {e}^c}
\newc{\superq}{\hat Q}
\newc{\superu}{\hat U}
\newc{\superd}{\hat D}
\newc{\superuc}{\hat {u}^c}
\newc{\superdc}{\hat {d}^c}

\newc{\tildeQ}{\widetilde Q}
\newc{\tildeU}{\widetilde U}
\newc{\tildeD}{\widetilde D}
\newc{\tildeL}{\widetilde L}
\newc{\tildeuc}{\widetilde {u}^c}
\newc{\tildedc}{\widetilde {d}^c}
\newc{\tildeec}{\widetilde {e}^c}
\newc{\tildenu}{\widetilde \nu}

\newc{\Lsoft}{{\cal L}_{soft}}
\newc{\hinot}{\widetilde H^0_2}
\newc{\hinob}{\widetilde H^0_1}
\newc{\bino}{\widetilde B^0}
\newc{\wino}{{\widetilde W^0_3}}
\newc{\gluino}{{\tilde g}}
\newc{\mgluino}{m_{\gluino}}
\newc{\msl}{m_{\tilde l}}
\newc{\msq}{m_{\tilde q}}
\newc{\msf}{m_{\tilde f}}
\newc{\mHu}{m_{H_u}}
\newc{\mHd}{m_{H_d}}
\newc{\mz}{m_Z}
\newc{\mw}{m_W}
\newc{\msel}{m_{\tilde e}}

\begin{document}

\title{Minimal supersymmetric standard model with gauge mediated 
supersymmetry breaking and neutrinoless double beta decay}

\author{Marek Góźdź}
\email{mgozdz@neuron.umcs.lublin.pl}
\author{Wiesław A. Kamiński} 
\email{kaminski@neuron.umcs.lublin.pl}
\author{Andrzej Wodecki}

\affiliation{Department of Theoretical Physics, Maria Curie--Skłodowska 
University, Lublin, Poland} 

\begin{abstract}
The Minimal Supersymmetric Standard Model with gauge mediated
supersymmetry breaking and trilinear $R$--parity violation is applied to
the description of neutrinoless double beta decay. A detailed study of
limits on the parameter space coming from the $B \to X_s \gamma$
processes by using the recent CLEO results is performed. The importance
of two--nucleon and pion-exchange realizations of $0\nu\beta\beta$ decay
together with gluino and neutralino contributions to this process are
addressed. We have deduced new limits on the trilinear $R$--parity breaking
parameter $\lambda_{111}'$ from the non-observability of
$0\nu\beta\beta$ in several medium and heavy open--shell nuclei for
different gauge mediated breaking scenarios. In general, they are
stronger than those known from other analyses. Also some studies with
respect to the future $0\nu\beta\beta$ projects are presented.
\end{abstract}


\maketitle

\section{Introduction}

During recent years, a lot of work has been devoted to test the Standard
Model (SM) of elementary particles. The best tested are interactions
between gauge bosons and matter, and in this sector the SM description
turns out to be very accurate. Other sectors, however, has been checked
to much less degree. Among them are self-interactions of gauge bosons as
well as the Higgs sector, which plays important role for completeness of
the model and in many aspects of symmetry breaking. Also many
shortcomings of SM, like the big number of free parameters, unresolved
question of mass hierarchy, the problem of massive neutrinos and their
oscillations, may call for more desirable description of Nature.

In the matter of fact, a number of various models reaching beyond SM's
orthodoxy were proposed. One of the most promising candidate is the
supersymmetric extention of SM called Minimal Supersymmetric Standard
Model (MSSM). It is based on the concept of supersymmetry (SUSY) and,
despite the lack of direct experimental evidence at the moment, is
supported by many theoretical arguments accompanied with hope, that SUSY
is the relevant description of our world above 1 TeV scale. One of the
main facts supporting MSSM is that incorporating SUSY in SM causes all
the gauge couplings unify at some scale $m_{GUT} \sim 10^{16}$ GeV. As
is well known, extrapolations of data from the LEP measurements suggest
such behaviour. However, SUSY particles have not been observed in
experiments, so supersymmetry has to be broken in low energy regime. The
issue how this breaking is realised, is the least understood question of
the theory.

The most widely studied version of SUSY conserves the so-called
$R$--parity. The $R$--parity is a multiplicativ quantity defined as $R=
(-1)^{2S+3B+L}$, where $B$ and $L$ are the baryon and lepton numbers,
and $S$ is the spin of corresponding particle. As a consequence,
processes which do violate lepton or baryon number are strictly
forbidden unless the symmetry is broken. Moreover, SUSY particles are
pair produced and the lightest one is stable.

The origin of $R$--parity conservation is not based on any fundamental
principle, so this property of MSSM is an {\it ad hoc} hypothesis and
therefore some extentions of the model, allowing the violation of
$R$--parity, were discussed in literature. These modifications can be
classified either as explicitely $R$--parity broken MSSM ($\Rslash$MSSM)
approaches \cite{hal84} or as formalisms with spontaneous breaking of
this symmetry \cite{aul83}. In the first class of models the
R--violating interactions are consistent with both gauge invariance and
SUSY \cite{bar89} while the second ones provide the simplest way for
$R$--parity violating effects conserving at low energy the baryon number
\cite{rom98}. The explicit $R$--parity breaking leads to well
defined phenomenological consequences, but due to large number of free
parameters involved, such theory has only marginal predictive power. In
contrast, the spontaneous breaking has many virtues added, like the
important possibility of dynamical origin of the $R$--parity breaking
\cite{rom98}.

Theories of gauge mediated supersymmetry breaking (GMSB) belong to the
second kind of approaches and have recently attracted a great deal of
attention. They are highly predictive, offer a natural solution to the
flavour problem and contain much less free parameters compared to MSSM
with SUSY breaking mediated by gravitational interaction
\cite{gmth1,gmth2,gmth3,gmth4,gmphen1,gmphen2,martin,haberkane}. In GMSB
models supersymmetry breaking is transmitted to the superpartners of
quarks, leptons and gauge bosons via the usual $SU(3) \times SU(2)
\times U(1)$ gauge interactions and occurs at the scale $M_{SUSY}\sim
10^5$ GeV. Gauginos and sfermions acquire their masses through
interactions with the messenger sector at one-- and two--loop levels
respectively, resulting in different phenomenology of the low--energy
world from the MSSM one. In these models flavour-diagonal sfermions mass
matrices are induced in a rather low energy scale, and therefore they
supply us with a very natural mechanism of suppressing unobserved in
experiments flavour changing neutral currents (FCNC). Moreover, since
the soft masses arise as gauge charges squared, the sizeable hierarchy
proportional to the gauge quantum numbers appears among the superpartner
masses. In this light, recently renewed interest in GMSB
\cite{gmphen1,gmphen2} is understandable.

The $R$--parity in MSSM can be explicitely violated by the presence of
bilinear \cite{valle} and trilinear \cite{rbreaking} terms in the
superpotential. The trilinear terms lead to lepton number and flavour
violation, while the bilinear terms generate non-zero vacuum expectation
values for sneutrino fields $\langle\tilde\nu_l \rangle$, causing
neutrino--neutralino and electron--chargino mixing. Thus, approaches
with lepton number violation can describe some low-energetic exotic
nuclear processes like the neutrinoless double beta decay
($0\nu\beta\beta$), known to be very sensitive to some of the
$R$--parity violating interactions \cite{koval1}. Using experimental
data about these processes, e.g. bounds on the half-life of neutrinoless
double beta decay, one can establish stringent limits on the $R$--parity
breaking SUSY \cite{koval1,koval2,wodecki,wodprc99,wodprd99}.

Supersymmetric models with $R$--parity non-conservation have been
extensively discussed in the last decade (see
e.g. \cite{rbreaking,valle}), and were also used for the study of
$0\nu\beta\beta$
\cite{koval1,koval2,wodecki,wodprc99,wodprd99,fae97}. The older
calculations were concentrated on the conventional two--nucleon mode of
$0\nu\beta\beta$, in which direct interaction between quarks of the two
decaying neutrons causes the process \cite{koval1,wodecki}. Recently the
dominance of pion exchange mode based on the double--pion exchange
between the decaying neutrons over the two--nucleon one was proved
\cite{fae97,fae98,wodprd99}.

Motivated by the forementioned features of GMSB models, in this paper we
study the $R$--parity breaking phenomenology of MSSM, and use the
neutrinoless double beta decay for deducing limits on some non-standard
physics parameters. In the previous studies such estimates were
performed in the framework of $\Rslash$MSSM with supergravity mediated
SUSY breaking by means of GUT constraints \cite{wodecki,wodprd99} or
additional assumptions relating sfermions and gauginos masses
\cite{koval1,koval2}. We will show that one can find quantitatively new
constraints \cite{wodprc99} within GMSB models. In this paper we study
this problem using up-to-date experimental data from CLEO collaboration
\cite{CLEO2001} for $0\nu\beta\beta$. As previously, we limit our
attention to the trilinear terms only, leaving complete treatment of
bilinear and trilinear $R$--parity violating terms in GMSB for
subsequent paper.

For reliable extraction of the limits on $R$--parity breaking constant
$\lambda _{111}'$ from the best presently available experimental lower
limit on the half-life of $0\nu\beta\beta$, it is necessary to determine
other SUSY parameters, e.g. masses of SUSY particles, within a proper
SUSY scenario, and to evaluate corresponding nuclear matrix
elements. Because at present the Renormalized Quasiparticle Random Phase
Approximation (RQRPA) \cite{toi95, SSF96}, which takes to some extent
the Pauli exclusion principle into account, is the main method commonly
used in calculations of the $0\nu\beta\beta$ nuclear matrix elements
\cite{fae98}, we used it also in our work.

Our article is organized as follows. In Section II the necessary theory
is developed. We also discuss to some extent the gauge mediated
supersymmetry breaking mechanism of the neutrinoless double beta decay.
Section III contains the results and analysis of constraints imposed on
supersymmetric parametres by non-observation of $0\nu\beta\beta$ in
Germanium $^{76}$Ge, for which the best experimental limit on the
half-life is known. In this part we also demonstrate differences between
the neutralino and gluino mechanisms in the neutrinoless double beta
decay. Finally, summary and concluding remarks can be found in Section
IV.

\section{Formalism}

\subsection{$R$--parity violation in MSSM}

In this section we briefly outline main features of MSSM and its
violation machanism. Both in the supergravity and in GMSB, the
$R$--parity can be explicitely violated by the bilinear \cite{valle} and
trilinear \cite{rbreaking} terms incorporated into the
superpotential. Bilinear terms generate non-zero vacuum expectation
values for the sneutrino fields $\langle \tilde \nu_L \rangle$, causing
neutrino--neutralino and electron--chargino mixing. Trilinear terms lead
to the lepton number and flavour violation. Above features make
$\Rslash$MSSM models appropriate for the description of
$0\nu\beta\beta$. Because this process is known to be very sensitive to
supersymmetric and $R$--parity breaking parameters, data from the
nowadays double beta experiments allow to establish stringent limits on
$\Rslash$MSSM physics
\cite{koval1,koval2,wodecki,fae97,fae98,wodprd99,wodprc99,hir99}.

The complete superpotential $W$ of the model can be written in the form
\begin{equation}      
  W = W_{0} + W_{\Rslash},
\label{eq:1} 
\end{equation}
where
\begin{eqnarray}
  W_0 = h^U_{ij} \superq_i\superhu\superuc_j + h^D_{ij}
          \superq_i\superhd\superdc_j &\nonumber \\ 
      + h^E_{ij}
	  \superl_i\superhd\superec_j + \mu \superhd\superhu &
\label{eq:2}
\end{eqnarray}
and
\begin{eqnarray}
  W_{\Rslash} = \lambda_{ijk} \superl_i\superl_j\superuc_k
           + \lambda_{ijk}'  \superl_i\superq_j\superdc_k   & \nonumber \\
	   + \lambda_{ijk}''  \superuc_i\superdc_j\superdc_k
	   + \mu_j\superl_j\superhu &
\label{eq:3}
\end{eqnarray}
are the $R$--parity conserving and $R$--parity breaking parts,
respectively. Here $\superq$ and $\superl$ denote the quark and lepton
$SU(2)$ doublet superfields, $\superuc$, $\superdc$, and $\superec$ the
corresponding $SU(2)$ singlets and $\superhu$, $\superhd$ are the Higgs
superfields. In the $R$--parity breaking part (\ref{eq:3}), the two
first terms are lepton number violating while the third violates the
baryon number conservation. The presence of these terms simultaneously
would cause unsuppressed proton decay and therefore we follow the
usual way and simply set $\lambda_{ijk} = \lambda_{ijk}'' = 0$ in
order to avoid such possibility.

In the low energy world supersymmetry is obviously broken and usualy one
supplies the theory with the ``soft'' breaking terms, being another source
of $R$--parity violation:
\begin{eqnarray}
&-&\Lsoft \nonumber \\
&=&\left(A^U_{ij} \tildeQ_i H_u \tildeuc_j
          + A^D_{ij} \tildeQ_i H_d \tildedc_j
	  + A^E_{ij} \tildeL_i H_d \tildeec_j + \hc \right)  \nonumber \\
&+& B\mu \left(H_d H_u + \hc \right)
    + m_{H_d}^2\vert H_d\vert^2 + m_{H_u}^2\vert H_u\vert^2 \nonumber\\
&+& m_{\tildeL}^2\vert \tildeL\vert^2 + m_{\tildeec}^2\vert
      \tildeec\vert^2 + m_{\tildeQ}^2\vert \tildeQ\vert^2 + 
       m_{\tildeuc}^2\vert \tildeuc\vert^2 + m_{\tildedc}^2\vert 
                                          \tildedc \vert^2 \nonumber \\
&+& \left(\frac 12 M_1 \bar{\psi}_B\psi_B + \frac 12
	M_2\bar{\psi}^a_W \psi^a_W + \frac 12 \mgluino\bar{\psi}^a_g
                                                 \psi^a_g + \hc \right)
\nonumber \\
\label{eq:4}
\end{eqnarray}
and
\begin{eqnarray}
-{\Lsoft}^{\Rslash} &=&
      {\tilde \lambda}_{ijk} \tildeL_i\tildeL_j{\tilde u}^c_k
    + {\tilde \lambda_{ijk}}'  \tildeL_i\tildeQ_j{\tilde d}^c_k
      \nonumber \\
&+& {\tilde \lambda_{ijk}}'' {\tilde u}^c_i{\tilde d}^c_j{\tilde d}^c_k 
    + {\tilde\mu}^2_{2j}\tildeL_j\superhu 
    +  {\tilde\mu}^2_{1j}\tildeL_j\superhd.
\label{eq:5}
\end{eqnarray} 
Here, fields with tilde denote the scalar partners of quark and
lepton fields, while $\psi_i$ are the spin-$1 \over 2$ partners of
gauge bosons.

To describe $0\nu\beta\beta$ process within supersymmetric models one
needs an explicit form of the appropriate Lagrangian. It can be
obtained using the standard procedure of extracting Lagrangian from
superpotential $W_{\Rslash}$. After some computation one gets
\begin{eqnarray}
  {\cal L}_{\lambda'_{111}} =
  -\lambda'_{111} \bigg[({\bar{u}_{L}},{\bar{d}_{L}})
     \left(
         \begin {array}{c}
           e_{R}^{c} \\ -\nu_{R}^{c}
         \end{array}
     \right)
  \tilde{d}_{R}^{\ast} &
\nonumber \\  
 +({\bar{e_{L}}},{\bar{\nu_{L}}}) d_{R}
     \left(
         \begin {array}{c}
           \tilde{u}^{\ast}_{L} \\ -\tilde{d}^{\ast}_{L}
         \end{array}
     \right) &
\\
 + ({\bar{u_{L}}},{\bar{d_{L}}}) d_{R}
     \left(
         \begin {array}{c}
           \tilde{e}^{\ast}_{L} \\ -\tilde{\nu}^{\ast}_{L}
         \end{array}
     \right)
 + \hc \bigg]&. \nonumber
\end{eqnarray}
Applying the formalism described in details in e.g. 
\cite{koval1,wodprd99}, one ends up with the effective Lagrangian:
\begin{eqnarray}
 {\cal L}^{\Delta L_e =2}_{eff}\ &=&
 \frac{G_F^2}{2 m_{_p}}~ \bar e (1 + \gamma_5) e^{\bf c} \nonumber \\
&\times& \left[\eta_{PS}~J_{PS}J_{PS} 
   - \frac{1}{4} \eta_T\   J_T^{\mu\nu} J_{T \mu\nu} \right],
\label{susy.2}
\end{eqnarray}
where the color singlet hadronic currents are $J_{PS} = {\bar
u}^{\alpha} (1+\gamma_5) d_{\alpha}$, $J_T^{\mu \nu} = {\bar u}^{\alpha}
\sigma^{\mu \nu} (1 + \gamma_5) d_{\alpha}$, with $\alpha$ being the
color index and $\sigma^{\mu \nu} = (i/2)[\gamma^\mu ,\gamma^\nu ]$.
The effective lepton number violating parameters $\eta_{PS}$ and
$\eta_{T}$ in Eq. (\ref{susy.2}) accumulate fundamental parameters of
MSSM and their explicit forms, obtained with a proper treatment of the
colour currents in the Lagrangian, can be found in Ref. \cite{wodprd99}.
These parameters are rather complicated functions of supersymmetric
masses and coupling constants which are, in general, free quantities
limited by experimental data or theoretical considerations only. In the
next section we describe our procedure how to obtain values of most of
them in the GMSB MSSM model.

\subsection{GMSB MSSM and procedure for finding supersymmetric parameters}

Supersymmetry breaking in GMSB models occurs in the so-called hidden (or
secluded) sector. It is a well known fact, that the detailed structure of
this sector does not change the phenomenology of low energy world. In
our approach we assumed that the secluded sector consists of a gauge
singlet superfield $\hat S$, whose lowest $S$ and $F$ components acquire
vacuum expectation values (vev).

Supersymmetry breaking is communicated to the visible world via the
so-called messenger sector. The interaction among superfields of the
secluded and messenger sectors is described by the superpotential
\begin{equation}
  W = \lambda_i \hat{S} \Phi_i {\overline \Phi_i}.
\label{gm3}
\end{equation}
where $\Phi_i$ and $\overline \Phi_i$ denote appropriate messenger
superfields. Because of non-zero vev of lowest $S$ and $F$ components of
superfield $\hat S$, fermionic components of messenger superfields gain
Dirac masses $M_i = \lambda_i S$ and determine in this way the messenger
scale $M$. Simultaneously mass matrices of their scalar superpartners
\begin{equation}
 \left( \begin{array}{cc}
   |\lambda_i S|^2 & \lambda_i F \\
   \lambda_i^* F^* & |\lambda_i S|^2 
 \end{array} \right) 
\label{gm4}
\end{equation}
have eigenvalues $|\lambda_i S|^2 \pm |\lambda_i F|$.

It is easy to see that vev of $S$ generates masses for fermionic and
bosonic components of messenger superfields, while vev of $F$ destroys
degeneration of these masses, which results in supersymmetry
breaking. Defining $F_i \equiv \lambda_i F$ one can introduce a new
parameter $\Lambda_i \equiv F_i/S$ measuring the fermion--boson mass
splitting:
\begin{eqnarray}
  m_f &=& M_i, \nonumber \\
  m_b &=& M_i \sqrt{1 \pm \frac{\Lambda_i}{M_i}}.
\end{eqnarray}
Parameter $\Lambda$ and messenger scale $M$ are in the following treated
as free parameters of the model.

Messenger superfields transmit SUSY breaking to the visible sector.  It
is realized through loops containing insertions of $S$ and results in 
gaugino and scalar masses at $M$ scale:
\begin{eqnarray}
  M_{\tilde \lambda_i}(M) &=& k_i \frac{\alpha_i(M)}{4 \pi}
\Lambda_G,  \\
\label{gm5}
  m^2_{\tilde f}(M) &=& 2 \sum_{i = 1}^{3} C_i^{\tilde f}k_i
  \left(\frac{\alpha_i(M)}{4 \pi}\right)^2\Lambda_S^2,
\label{gm6}
\end{eqnarray}
where  $i=1,2,3$ is the gauge group index, and
\begin{eqnarray}
  \Lambda_G &=&\sum_{k = 1}^{N_g}n_k\frac{F_k}{M_k}
  g\left(\frac{F_k}{M_k^2}\right), 
\label{gm71} \\
  \Lambda_S^2 &=& \sum_{k = 1}^{N_g}n_k\frac{F_k}{M_k^2}
  f\left(\frac{F_k}{M_k^2}\right), 
\label{gm7}
\end{eqnarray}
with $k$ being the flavour index. In Eqs. (\ref{gm71}) and
(\ref{gm7}) $n_k$ is the doubled Dynkin index of the messenger
superfield representation with flavour $k$. Coefficients $C_i^{\tilde
f}$ are the quadratic Casimir operators of sfermions. For $d$--dimensional
representation of SU(d) their eigenvalues are $C = (d^2 - 1)/2d$. In
the case of $U(1)$ group $C = Y^2 = (Q - T_3)^2$. It follows that
coefficients $k_i$ are equal to $5/3$, $1$ and $1$, for $SU(3)$,
$SU(2)$, and $U(1)$ respectively. The normalization here is conventional
and assures that all $k_i\alpha_i$ meet at the GUT scale. Finally, the
functions $f$ and $g$ have the following forms:
\begin{equation}
  g(x) = \frac{1}{x^2}[(1+x)\log(1+x)] + (x \to -x),
\label{ggm}
\end{equation}
\begin{eqnarray}
  f(x) &=& \frac{1 + x}{x^2}
  \bigg[\log(1 + x) - 2 Li_2\left(\frac{x}{1 + x}\right) \nonumber \\
    &+& {1 \over 2}Li_2\left(\frac{2x}{1 + x}\right)\bigg] + (x \to -x). 
\label{fgm}
\end{eqnarray}

The minimal model of GMSB considered in this paper contains only one
messenger field flavour. Thus, dropping flavour indices, one can write
\begin{eqnarray}
 & M_{\tilde \lambda_i}(M) = N k_i \frac{\alpha_i(M)}{4 \pi}
  \Lambda g\left(\frac{\Lambda}{M}\right), 
\label{gm10}\\
 & m^2_{\tilde f}(M) = 2 N \sum_{i = 1}^{3} C_i^{\tilde f}k_i
  \left(\frac{\alpha_i(M)}{4\pi}\right)^2\Lambda^2
  f\left(\frac{\Lambda}{M}\right) \cdot {\bf 1}, \nonumber \\
\label{gm11}
\end{eqnarray}
where $C_1^{\tilde f} = Y^2, C_2^{\tilde f} = 3/4$ for $SU(2)_L$
doublets and 0 for singlets, $C_3^{\tilde f}$ is equal to $4/3$ for
$SU(3)_C$ triplets and 0 for singlets. In (\ref{gm11}) ${\bf 1}$ denotes
the unit matrix in generation space and guarantees the lack of flavour
mixing in soft breaking mass matrices at messenger scale. $N$, the
so-called generation index, is given by $N = \sum_{i = 1}^{N_g}n_i$,
where $N_g$ means the total number of generations. In this paper we
study two cases:
\begin{enumerate}
\item a single flavour of $5 + \overline 5$ representation
of $SU(5)$, with $SU(2)_L$ doublets ($l$ and $\tilde l$), and $SU(3)$
triplets ($q$ and $\tilde q$);
\item a single flavour of both representations $5 + \overline 5$
and $10 + \overline{10}$ of $SU(5)$ group.
\end{enumerate}
In case 1. $N$ is equal to 1, while in case 2. $N = 1 + 3 = 4$,
because for $10 + \overline{10}$ representation of $SU(5)$ the doubled
Dynkin index is 3.

\subsection{Renormalization Group Equations and parameters
  determination}

The evolution of all running parameters is realized using
Renormalization Group Equations (RGE). The formulae (\ref{gm10}) and
(\ref{gm11}) may therefore serve as boundary conditions for evolution of
soft parameters at the electroweak scale. Our procedure resulting in low
energy spectrum of SUGRA and GMSB MSSM models and its application to the
description of $0\nu\beta\beta$ decay can be found in our previous
papers \cite{wodprc99,wodprd99}, so here we only sketch its most
important features.

At the beginning, one evolves all gauge and Yukawa couplings for three
generations up to the messenger scale $M$. We use the one--loop Standard
Model RGE \cite{drtjones} below the mass threshold, where SUSY particles
start to contribute, and MSSM RGE \cite{martinvaughn} above that
scale. We admit not to use the full set of RGE appropriate for the
$\Rslash$MSSM model \cite{car96,allanach}. The influence of $R$--parity
breaking constants on other quantities running from the messenger to the
electroweak scale is marginal due to the smallness of $\lambda$'s. In
our case the two--loop corrections can also be safely neglected (for a
discussion of this problem see \cite{mg-art1}). Initially, scale
$M_{SUSY}$ is taken to be equal to 1 TeV, but it is dynamically modified
during running of relevant masses. In the next step we construct the
gaugino and sfermion soft mass matrices using Eqs. (\ref{gm10}) and
(\ref{gm11}), and perform RGE evolution of all the quantities back to
$\mz$ scale. During this run, $m^2_{H_u}$ reaches negative value causing
dynamical electroweak symmetry breaking (EWSB). It is well known, that
proper treatment of this mechanism needs minimizing of the full one--loop
Higgs effective potential \cite{finitehiggs}. On the other hand,
appropriate corrections contain functions of particle mass eigenstates
generated by EWSB mixing. Thus, as the first approximation, we minimize
the tree-level Higgs potential parameters $\mu$ and $B\mu$ which are
crucial for further analysis.
\begin{figure}
\includegraphics{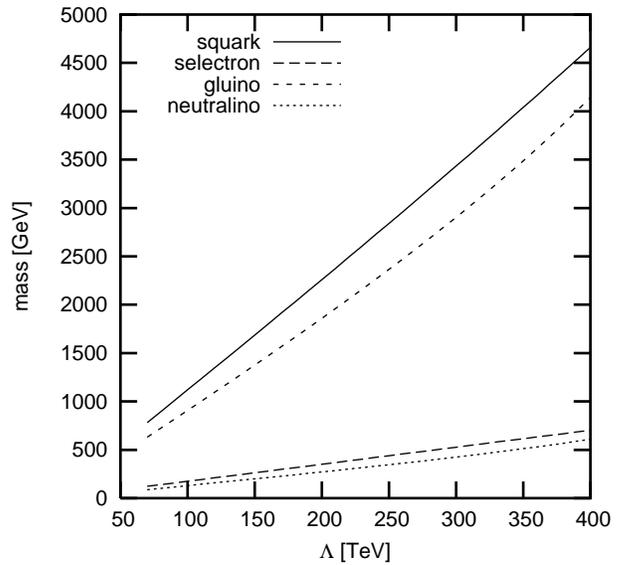}
\caption{\label{fig1} Sample RGE evolution of sparticle masses in GMSB
  MSSM and their dependence on $\Lambda$.}
\end{figure}

Having all needed mass parameters at electroweak scale, one can evolve
all other quantities to some scale $Q_{min}$, which is optimal for
minimization of the one--loop corrected Higgs potential. At this scale,
defined as the geometric mean of stop masses, minimization procedure
results in $\mu$ and $B\mu$ values. Next, all the quantities are running
back to $\mz$ scale. Iterating this procedure one obtains stable values
of $\mu$ and $B\mu$ and then low energy spectrum for the considered
model. Only four parameters: $\Lambda$, $M$, $\tan\beta \equiv v_u/v_d$
and $\rm sgn(\mu)$ remain free. The quantities $v_u$ and $v_d$ are
vev's of $\superhu$ and $\superhd$, respectively.

In Fig. \ref{fig1} a sample evolution of sparticle masses versus the
$\Lambda$ parameter is shown. Other parameters were $\tan\beta =3$, $M =
500$ TeV, $N=1$. One sees that the masses of squark and gluino depend
heavily on $\Lambda$, while in the case of selectron and neutralino this
dependence is much weaker.

\subsection{Restrictions on low energy spectrum}

\begin{figure}
\includegraphics{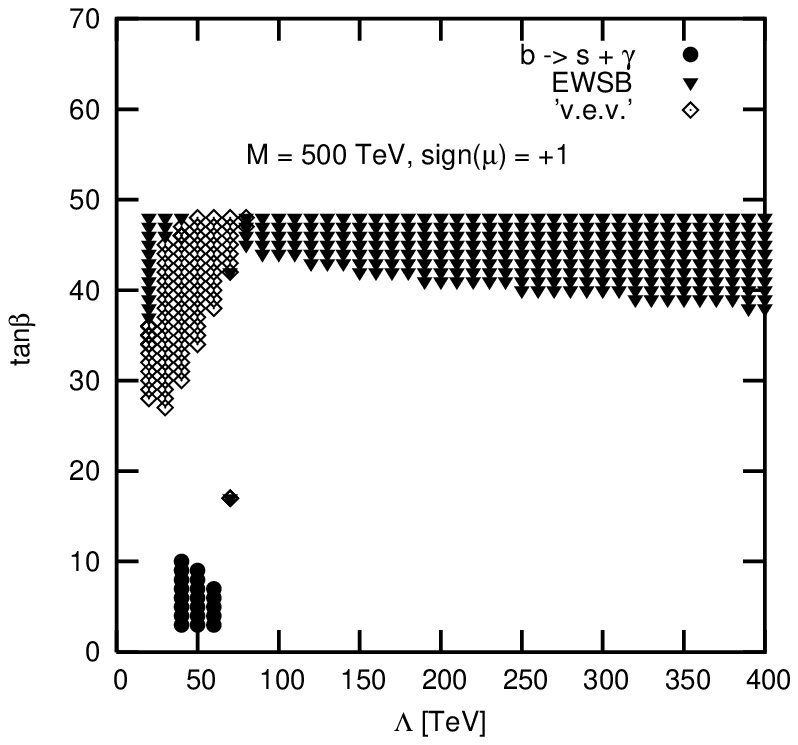}
\includegraphics{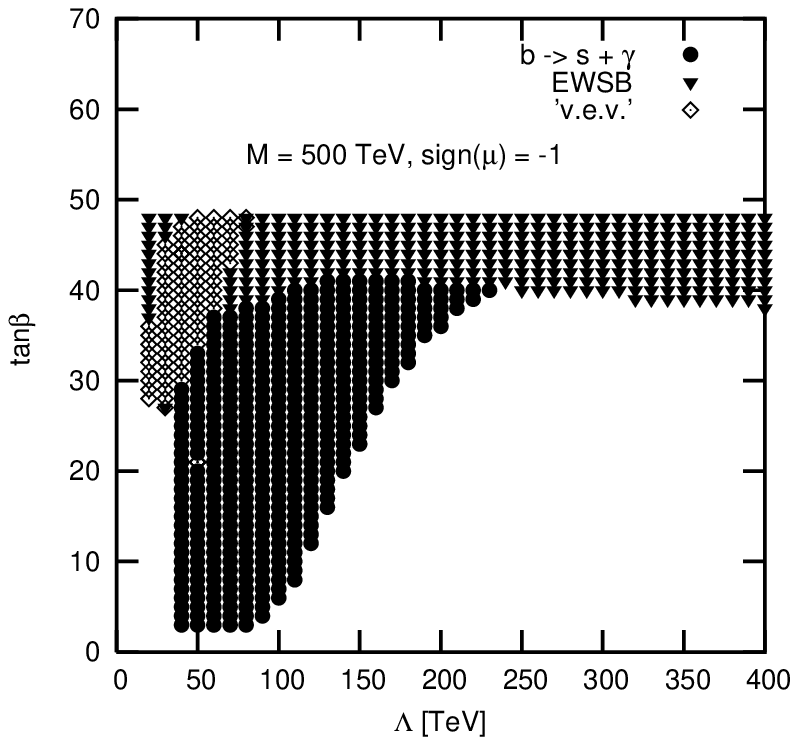}
\caption{\label{fig2} Constraints on GMSB parameter space.}
\end{figure}

To impose restrictions coming from the present theoretical assumptions and
phenomenological data on the resulting spectrum is a non-trivial problem.
We would like to obtain limits on physics beyond SM induced by $0\nu \beta
\beta$ experiments, consistent with constraints coming from: 
\begin{enumerate}
\item finite values of Yukawa couplings at the GUT scale;
\item proper treatment of electroweak symmetry breaking;
\item requirement of physically acceptable mass eigenvalues at low
energies;
\item flavour changing neutral currents (FCNC) phenomenology.
\end{enumerate}
Below we will briefly discuss these sources of additional constraints.

The first requirement comes from the RGE evolution procedure. It is well
known that running of the Yukawa couplings is sensitive to initial (i.e.
at the electroweak scale) values determined by $\tan\beta$. For very
small $\tan\beta$ $(< 1.8)$ the top Yukawa coupling may ``explode''
before reaching the GUT scale. It follows from the fact, that
$Y_{top}(\mz) \sim 1/\sin \beta$. Similarly, other couplings $Y_{b}$ and
$Y_\tau$ ``blow up'' before the GUT scale for $\tan\beta > 50$ because
they are proportional to $ 1/\cos \beta$ at electroweak scale. Such
behaviour of the Yukawa couplings limits the range of $\tan\beta$ to the
interval 2--50.

Another theoretical constraint is imposed by the EWSB mechanism. In
order to obtain a stable minimum of the scalar potential, the following
conditions must hold:
\begin{eqnarray}
  (\mu B)^2 &>& \left( \left|\mu\right|^2 + \mHu^2 \right)
  \left( \left|\mu\right|^2 + \mHd^2 \right), \nonumber \\
  2B\mu &<& 2 \left|\mu\right|^2 + \mHu^2 + \mHd^2.
\label{EWSB}
\end{eqnarray}
They are always checked in our procedure during RGE running, and points
which do not fulfill these conditions are rejected (see Fig. \ref{fig2},
points marked ``EWSB''). Next restriction comes from the requirement of
positive eigenvalues of mass matrices squared at the electroweak scale,
and allows to find combinations of free parameters providing the
negative (forbidden) eigenvalues marked in Fig. \ref{fig2} as
``v.e.v.''.

The most interesting set of constraints has its source in the FCNC
phenomenology. Such processes, strongly experimentaly suppressed, limit
upper values of different entries of the sfermion mass matrices at low
energies
(cf. \cite{bsgmssm,bsggmsb,bsgnlo,bsg_pokorscy,bsg_mssm_nlo,fcnc2000}).
Here we consider the $B \to X_s \gamma$ decay only. The effective
Hamiltonian for this process reads \cite{bsgmssm,bsg_pokorscy}
\begin{equation}
  H_{\rm eff} = - \frac{4G_F}{\sqrt{2}}K^*_{ts}K_{tb}
  \sum_{i = 1}^8C_i(\mu)P_i(\mu),
\label{e:20}
\end{equation}
where $K$ is the quark mixing matrix (CKM matrix) and $P_i$ are the
relevant operators taken from Ref. \cite{bsg_pokorscy}. Among the Wilson
coefficients $C_i(\mu)$ two: $C_7$ and $C_8$, are the most important for
the analysis of impact of the SM and MSSM interactions. (The leading
order and the next--to--leading order analysis of these interactions
were discussed in \cite{bsgmssm,bsg_mssm_nlo}.)  In order to costrain
the low energy spectrum of supersymmetric models using FCNC processes,
it is a common practice to define the parameter $R_7$, which measures the
extra (MSSM) contributions to the $B \to X_s \gamma$ decay:
\begin{equation}
  R_7 \equiv 1 + \frac{C_7^{(0)extra}(\mw)}{C_7^{(0)SM}(\mw)},
\label{eq:21}
\end{equation}
where the index $(0)$ stands for the leading order Wilson
coefficients and the superscript $extra$ indicates SUSY (charged Higgs,
chargino, neutralino and gluino) contributions. Explicit expressions
for $C_7^{(0)extra}$ and $C_7^{(0)SM}$ can be found e.g. in
Ref. \cite{bsgmssm}. Constraints on allowed values of $R_7$ are induced
from the present experimental limits on the branching ratio BR($B \to
X_s \gamma$) measured by CLEO collaboration \cite{CLEO2001}:
\begin{equation}
  {\rm BR}(B \to  X_s \gamma) = 
  (3.21 \pm 0.43_{stat} \pm 0.27_{syst})\times 10^{-4} .
\end{equation} 
The theoretical dependence of BR($B \to X_s \gamma$) on $R_7$ confronted
with such experimental data allows to make the following estimate:
\begin{equation}
  -6.6 < R_7 < -4.4 \qquad \hbox{or} \qquad 0.0 < R_7 < 1.3.
\label{eq:22}
\end{equation}
Using the above restriction, one can exclude certain values of
supersymmetric parameters, which result in the $R_7$ coefficient outside
the allowed region (\ref{eq:22}). In Fig. \ref{fig2} such points are
marked as ``b $\to$ s + $\gamma$''.

Looking on Fig. \ref{fig2} one sees that the constraints deduced from
FCNC phenomenology are very sensitive to the sign of $\mu$. The same
behaviour was also observed in SUGRA MSSM model (see
e.g. Ref. \cite{wodprd99}) and is mainly due to sensitivity of charged
Higgs and chargino contributions to $R_7$ on sign of the $\mu$
parameter.

\begin{figure}
\includegraphics{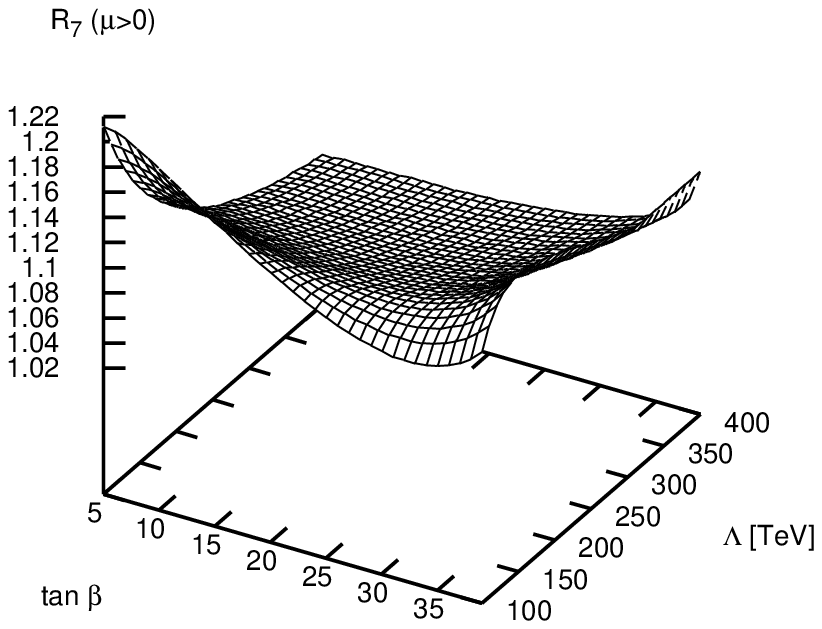}
\includegraphics{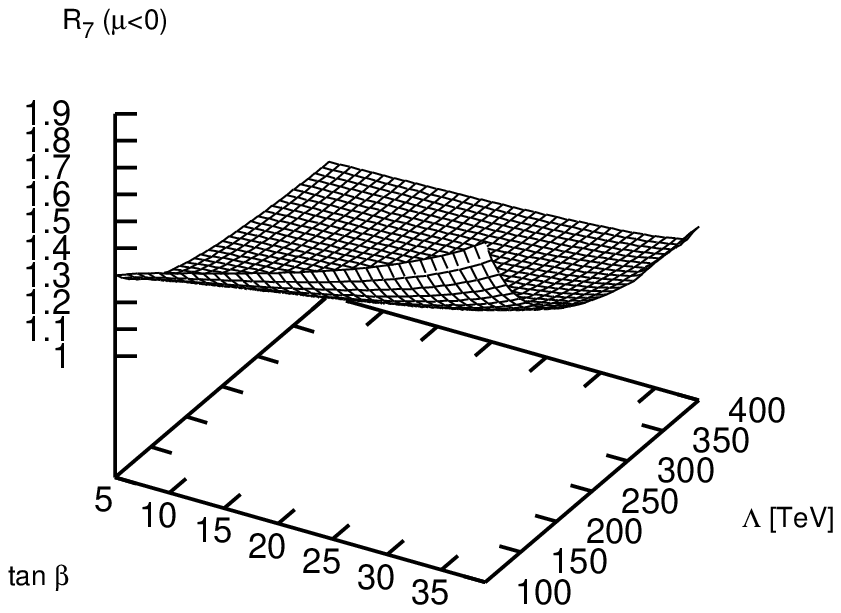}
\caption{\label{fig3} $R_7$ parameter in GMSB MSSM for both signs of
  $\mu$. The scan is performed over $\Lambda$ and $\tan\beta$, with $M =
  500$ TeV and $N$ = 1.}  
\end{figure}
%

\begin{figure*}
\includegraphics{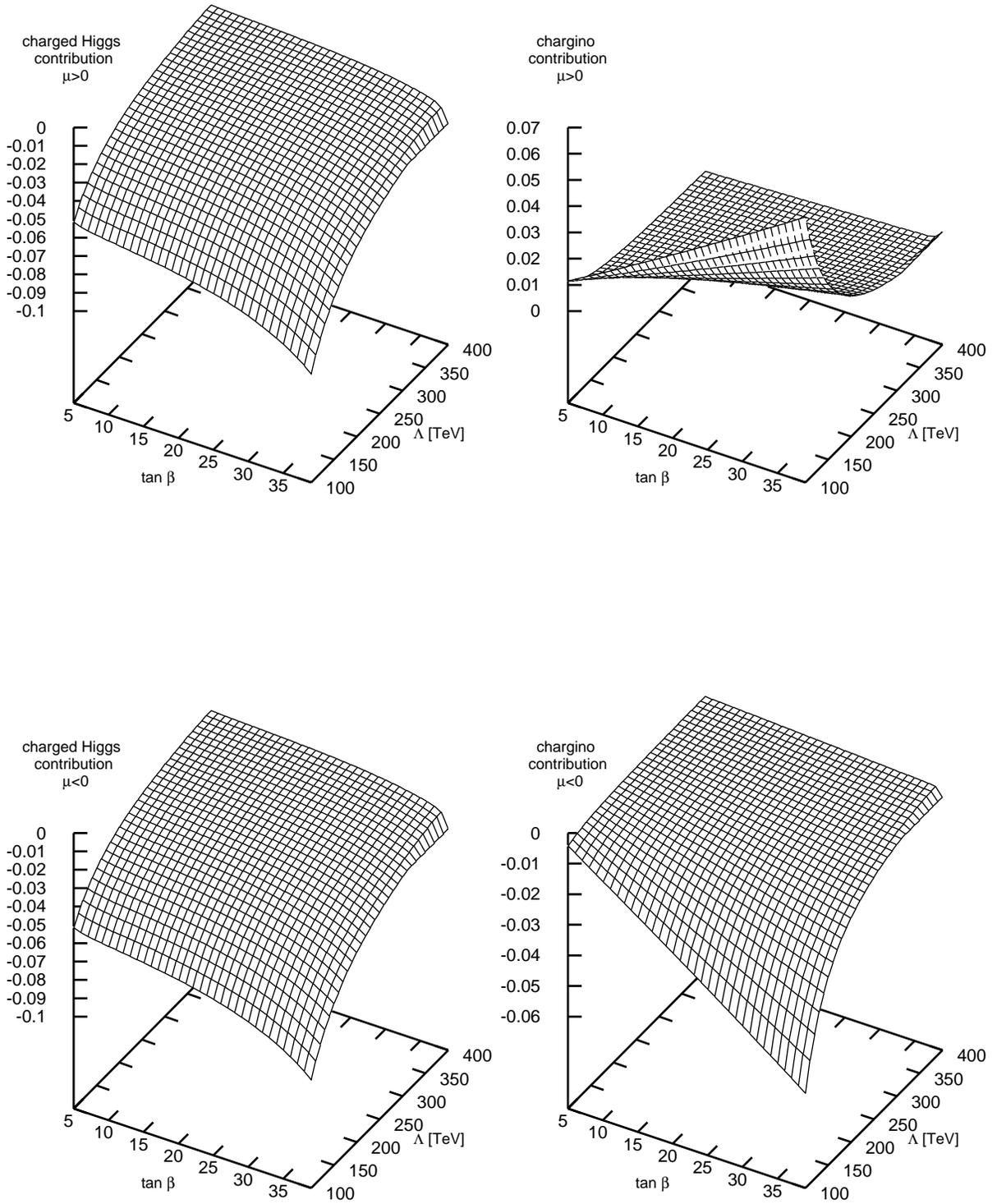}
\caption{\label{fig4} Contributions to $C_7^{(0)MSSM}$ in GMSB MSSM
  coming from charged Higgses and charginos for both signs of $\mu$. All
  parameters as in Fig. \ref{fig3}} 
\end{figure*}

The additional dependence of $R_7$ on both $\tan\beta$ and $\Lambda$
parameters is shown for positive and negative $\mu$ in
Fig. \ref{fig3}. In the case $\mu > 0$ the parameter $R_7$ grows up for
smaller values of $\tan\beta$ and behaves in opposite manner for $\mu <
0$. Moreover, in the latter case $R_7$ is, in general, bigger which
results in more stringent restrictions. More detailed analysis is
presented in Fig. \ref{fig4}, where the most important impacts to $R_7$
for different choices of $\tan\beta$, $\Lambda$, and $\rm sgn(\mu)$ are
explicitely shown. One can see that $\tan\beta$ and $\rm sgn(\mu)$ do
not influence the charged Higgs contribution significantly. Thus, a
crucial point in the analysis becomes chargino contribution. Contrary to
SUGRA MSSM case \cite{wodprd99} the magnitude of chargino influence on
$R_7$ is almost equal to the influence coming from charged Higgses. For
positive values of the $\mu$ coupling constant, the chargino impact
grows with increasing of $\tan\beta$, while for $\mu < 0$ case one
observes opposite behaviour. In this light, behaviour of the surfaces
shown in Fig. \ref{fig3} becomes clear.

\section{Neutrinoless double beta decay and limits on non-standard physics}

\begin{figure}
\includegraphics{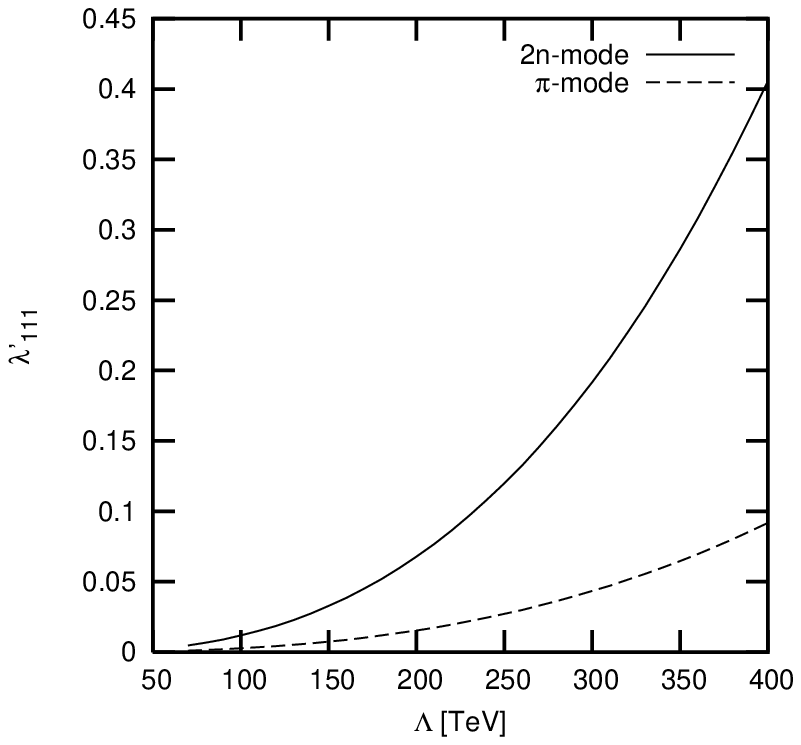} 
\includegraphics{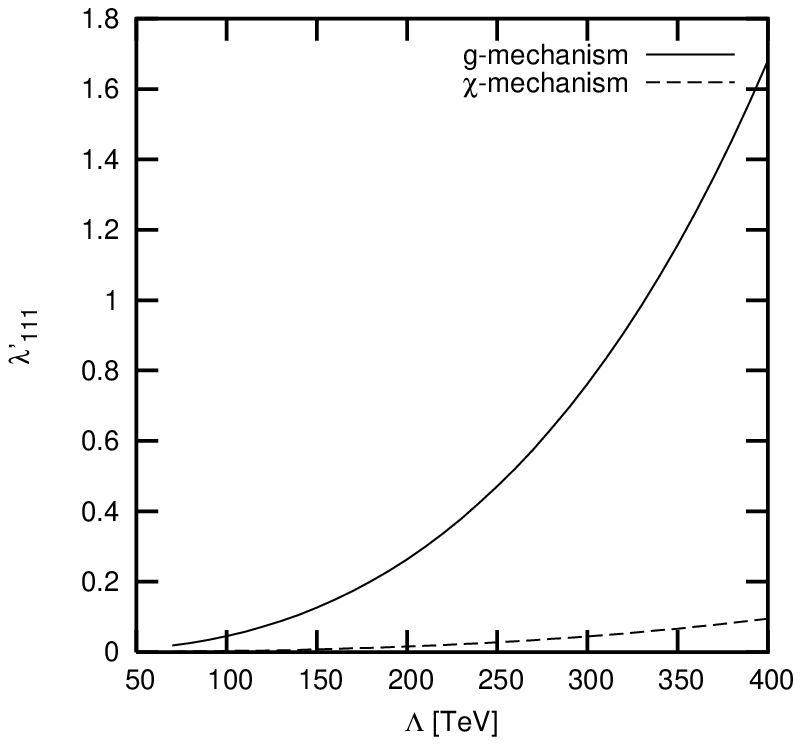}
\caption{\label{fig5} Contributions to $\lambda'_{111}$ coming from
  two-nucleon and pion-exchange modes of $0\nu\beta\beta$ in $^{76}$Ge
  as well as gluino and neutralino mechanism of SUSY breaking. All
  parameters like in Fig. \ref{fig3} except $\tan\beta = 3$ and $\mu>0$.}
\end{figure}

Restrictions imposed on the model by low energy spectrum allow for
reliable analysis of exotic nuclear processes, like the neutrinoless
double beta decay, and then for deduction of additional constraints
imposed on non-standard physics. In this paper we use experimental
information about non-observability of the $0\nu\beta\beta$ decay in
different nuclei to extract stringent limits on $R$--parity breaking.

The half-life of the process, taking into account all three possible
types of hadronization (2--nucleon, 1--pion and 2--pion
\cite{koval1,wodprd99}) reads:
\begin{eqnarray}
\big[ T_{1/2}^{0\nu} \big]^{-1}
&=&  G_{01} \bigg | \eta_{T} {\cal M}_{\tilde q}^{2N} 
+ \Big(\eta_{PS} - \eta_{T}\Big) {\cal M}_{\tilde f}^{2N} \nonumber \\
  &+& \frac{3}{8} \left(\eta_{T} + \frac{5}{8} \eta_{PS}\right)
  {\cal M}^{\pi N} \bigg |^2.
\label{eq:29}
\end{eqnarray}
In this equation ${\cal M}_{\tilde q}^{2N}, {\cal M}_{\tilde f}^{2N}$
and ${\cal M}^{\pi N}$ are matrix elements for the $2N$, $1\pi$ and
$2\pi$ channels, respectively. These matrix elements depend on
non-standard physics parameters, involved in description of the
neutrinoless double beta decay, and on nuclear structure details of
decaying nuclei. (The explicit forms of elements (\ref{eq:29}) can be
found, e.g., in \cite{koval1,wodprd99}.) Our procedure limits the number
of free parameters to $\Lambda$, $M$, $\tan\beta$, sign$(\mu)$, and $N$
only. As the loop diagrams with messenger fields do not affect the
A--terms considerably, we can equal the common soft SUSY breaking
parameter $A_0$ to 0 at the $M$--scale.

Following well established procedure, the nuclear matrix elements in
question were calculated within the proton--neutron Renormalized
Quasiparticle Random Phase Approximation (pn--RQRPA). This approach
incorporates the Pauli exclusion principle for fermion pairs
\cite{toi95,SSF96} and is suitable for studies of nuclear structure
aspects of various double beta decay channels in open shell systems.
Details of the method and its application to the double beta decay were
presented in a number of articles (see, e.g., \cite{gmphen1, wodprd99,
fae98}).

Having both supersymmetric spectrum and nuclear matrix elements, one can
extract from Eq. (\ref{eq:29}) constraints on $R$--parity breaking in GMSB
MSSM using experimental information from non-observability of the
neutrinoless double beta decay. Such approach is based on
comparison of the theoretically obtained half-life for this process, as
a function of some free non-standard parameters, with the experimental
upper limit for $T_{1/2}$ in the given nucleus.

We start with a presentation of constraints on $\lambda'_{111}$ coming
from different channels of $0\nu\beta\beta$. In Ref. \cite{fae97} the
problem of the pion mode has been discussed in details. In
Fig. \ref{fig5} the importance of pion--exchange mode is clearly
visible. The curve corresponding to the pion mode lies definitely below
the line corresponding to the nucleon channel, so the pion mode imposes
more stringent restrictions on the coupling constant. Also the role of
various mechanisms leading to SUSY breaking are presented. These data
were calculated for $^{76}$Ge nucleus, for which the best experimental
limits, coming from the IGEX collaboration, are known \cite{IGEX}. One
sees that the SUSY breaking mediated by neutralinos sets more severe
limits on $\lambda'_{111}$ than the gluino mechanism.

\begin{figure}
\includegraphics[width = 8.5 truecm, height = 17 truecm]{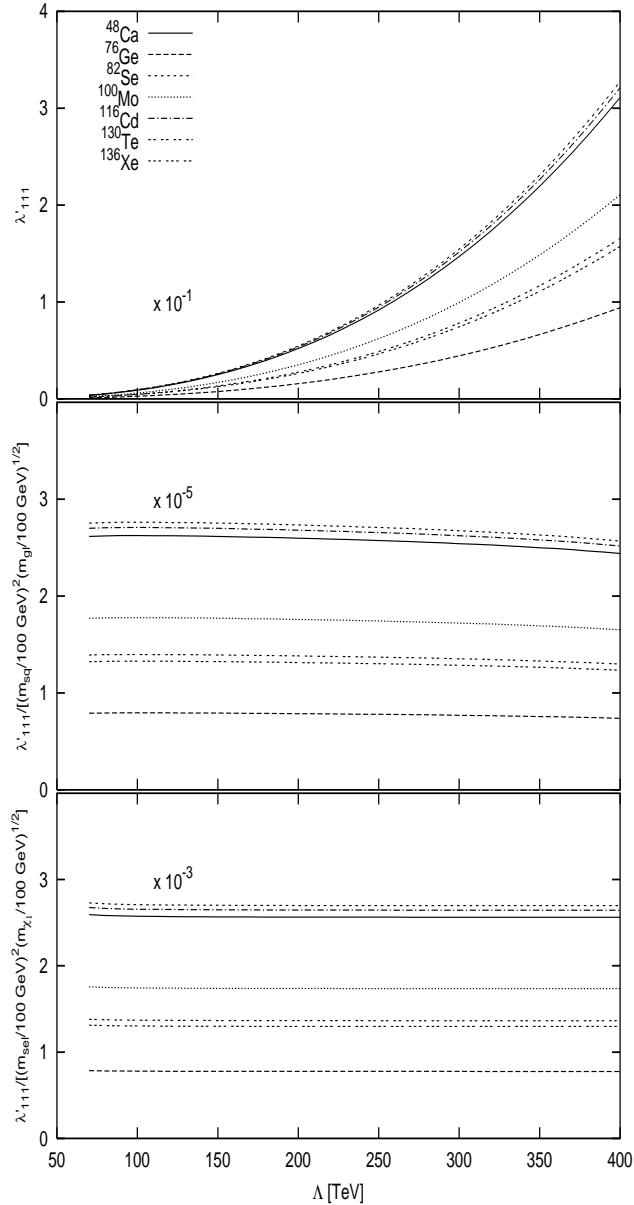}
\caption{\label{fig6} Limits in GMSB MSSM on various combinations of
  $\lambda_{111}'$ and masses of SUSY particles coming from experimental
  lower bounds on the half-life of $0\nu\beta\beta$ decay in different
  nuclei. The corresponding nuclear matrix elements have been calculated
  using pn-RQRPA method and the bag model. Other parameters as in
  Fig. \ref{fig3}} 
\end{figure}
\begin{figure}
\includegraphics{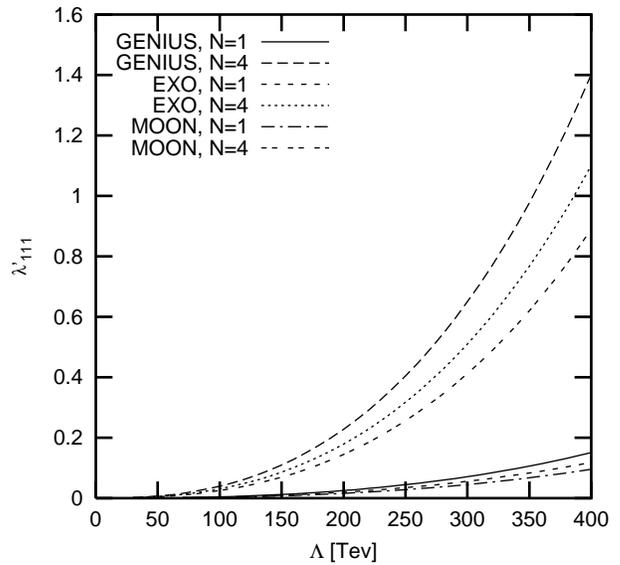}
\caption{\label{fig7} Limits on $\lambda_{111}'$ for different
  structures of messeneger sector (see text for details) for planned
  $0\nu\beta\beta$ experiments. All parameters as in Fig. \ref{fig5}}
\end{figure}

Further analysis is presented in Fig. \ref{fig6}. We have included most
of the nowadays known experimental data (see \cite{now-exp} and
references therein). The interesting thing is, that the combinations
$\lambda'_{111}/[(\msq/100\GeV)^2(\mgluino/100\GeV)^{\frac12}]$ and
$\lambda'_{111}/[(\msel/100\GeV)^2(m_{\chi1}/100\GeV)^{\frac12}]$ remain
nearly unchanged within wide range of $\Lambda$'s. This allows us to
estimate the lepton number violating constant:
\begin{equation}
\frac{\lambda'_{111}}{(\frac{\msq}{100 \GeV})^2 \sqrt{\frac{\mgluino}{100
      \GeV}}} < 2.75 \cdot 10^{-5}
\end{equation}
and 
\begin{equation}
\frac{\lambda'_{111}}{(\frac{\msel}{100 \GeV})^2 \sqrt{\frac{m_{\chi1}}{100
      \GeV}}} < 2.73 \cdot 10^{-3}.
\end{equation}
These results lower the allowed values in the first case by around 15\%
(cf. \cite{wodprc99}). 

We study also constraints coming from different GMSB scenarios in the
case of expected sensitivity of planned neutrinoless double beta decays.
Two different messenger sector structures are here taken into account:
the $5 + \overline{5}$ representation ($N = 1$) and both, $5 +
\overline{5}$ and $10 + \overline{10}$, representations ($N = 4$) of
$SU(5)$. We include parameters for three new experiments
\cite{future-exp, now-exp}. The GENIUS--MAJORANA-GEM project is expected
to reach sensitivity of $T_{1/2}\sim 2.3\cdot10^{28}$y for
$^{76}$Ge. The MOON experiment has $T_{1/2}\sim 1.3\cdot10^{28}$y and
investigates the $^{100}$Mo nuclei, and the EXO-XMASS experiment will be
sensitive to values of the half-life up to around $T_{1/2}\sim
2.2\cdot10^{28}$y for $^{136}$Xe. The relevant results are presented in
Fig. \ref{fig7}. It is worth noting that for $N=4$ the allowed values
for the lepton number violating constant are much higher. The most
promising results can be expected from the MOON project, which may set
the best constraints on the $R$-parity violating coupling constant.

\section{Conclusions}

We have presented an analysis of the current experimental state of
neutrinoless double beta decay in the language of gauge mediated Minimal
Supersymmetric Standard Model. Combining theoretical, phenomenological,
and experimental data we obtained a set of constraints on various
non-standard parameters. In particular, new upper limits for
$\lambda'_{111}/[(\msq/100\GeV)^2(\mgluino/100\GeV)^{\frac12}]$ and
$\lambda'_{111}/[(\msel/100\GeV)^2(m_{\chi1}/100\GeV)^{\frac12}]$ were
extracted. A detailed discussion of the Wilson coefficients, the SUSY
contributions to it and its dependence on the whole allowed range of
$\tan\beta$ and $\Lambda$ up to 400 TeV was presented. Also some
preliminary studies related to three new planned $0\nu\beta\beta$
experiments were performed.


\end{document}